\documentclass[a4paper,11pt]{article}
\pdfoutput=1 

\usepackage{jcappub} 

\usepackage[T1]{fontenc} 

\title{Non-slow-roll dynamics in $\alpha-$attractors}

\author[a,b]{K. Sravan Kumar}
\author[a,b]{J. Marto}
\author[a,b]{P. Vargas Moniz}
\author[c]{Suratna Das}

\affiliation[a~]{Departamento de F\'{\i}sica, Universidade da Beira Interior (UBI), \\ 6200, Covilh\~{a}, Portugal}
\affiliation[b~]{Centro de Matem\'atica e Aplica\c{c}\~oes da Universidade da Beira Interior (CMA-UBI), \\ 6200, Covilh\~{a}, Portugal}
\affiliation[c~]{Indian Institute of Technology, Kanpur 208016, India.}

\emailAdd{sravan@ubi.pt}
\emailAdd{jmarto@ubi.pt}
\emailAdd{pmoniz@ubi.pt}
\emailAdd{suratna@iitk.ac.in}

\abstract{
In this paper we consider the $\alpha-$attractor model and study
inflation under a non-slow-roll dynamics. More precisely, we follow the approach
recently proposed by Gong and Sasaki  \cite{Gong:2015ypa}
by means of assuming $N=N\left(\phi\right)$. Within this framework we obtain a family of functions describing the local shape of the potential during inflation. We study a specific model and find an inflationary scenario  predicting an  attractor at $n_{s}\approx0.967$ and $r\approx5.5\times10^{-4}$.
We further show that considering a non-slow-roll dynamics, the $\alpha-$attractor
model can be broaden to a wider class of models that remain compatible with value of $r<0.1$. We further explore the model parameter space with respect to large and small field inflation and conclude that the inflaton dynamics is connected to the $ \alpha- $ parameter, which is also related to the  K\"ahler manifold curvature in the supergravity (SUGRA) embedding of this model. We also comment on
the stabilization of the inflaton's trajectory.  }


\makeatother

\begin{document}
\maketitle 

\section{Introduction}

Inflationary cosmology has become an extremely convincing paradigm for
the early universe after the recent release of Planck data \cite{Ade:2015lrj,Ade:2015tva,Ade:2015ava}.
The unprecedented accuracy of the data is consistent with the LCDM, nearly
scale invariant and Gaussian nature of curvature perturbations. We
now have stringent bounds on the scalar spectral index, namely, $n_{s}=0.968\pm0.006$
at 95\% CL and also for the tensor to scalar ratio, which is severely
bounded, $r<0.1$ ($ r<0.07 $ considering recent results of BICEP2/Keck Array \cite{Array:2015xqh}). Given these circumstances, single field inflation
constitutes a scenario which is in full agreement with the data. However,
the nature of the inflaton is still elusive \cite{Martin:2013tda}. 

Among the broad variety of inflationary scenarios, the Starobinsky model, with a $R+R^{2}$ term, and the Higgs
inflation \cite{Starobinsky:1980te,Starobinsky:1983zz,Bezrukov:2007ep}
stand in a privileged region in the middle of the $\left(n_{s},\, r\right)$
plane \cite{Ade:2015lrj}. Moreover, the Starobinsky model prediction
is identified as a target spot for many inflationary
models, i.e,

\begin{equation}
n_{s}=1-\frac{2}{N}\quad r=\frac{12}{N^{2}}\,,\label{sweetspot}
\end{equation}

where $N$ is the $50-60$ number of e-foldings before the end of inflation. The requirement of $r\lesssim\mathcal{O}\left(10^{-2}\right)$ and
the nearly vanishing non-Gaussianities is compatible with inflationary models
embedded in a low energy effective field theories derived from a UV-completion
physics, such as supergravity (SUGRA) and string theory \cite{Linde:2014nna,Burgess:2013sla,Martin:2015dha}.

Since the first release of Planck 2013 data, these two scenarios (Starobinsky
model and Higgs inflation) started to attract a lot of attention, they
have been extensively studied and realized in the context of conformal
symmetries \cite{Kallosh:2013hoa,Kallosh:2013daa}, later generalized
as $\alpha-$ and non-minimal (or) $\xi-$attractors. In addition,
these models have been embedded in SUGRA through the
use of superconformal symmetries \cite{Kallosh:2013lkr,Kallosh:2013yoa,Kallosh:2013maa,Kallosh:2013pby,Kallosh:2014rga}.
Recently, $\alpha-$ attractor models were also realized by means of the inclusion of an auxiliary vector field for the Starobinsky model \cite{Ozkan:2015iva}.
These two classes of models have also, a posteriori, been unified as
cosmological attractor models (CAM) \cite{Galante:2014ifa,Cecotti:2014ipa,Roest:2013fha}.
By varying the parameters $\left(\alpha,\xi\right)$ in CAM, on the one hand leads to the Starobinsky attractor (\ref{sweetspot})
and on the other hand it also reproduces the chaotic inflation predictions with the $m^{2}\phi^{2}$
potential. In particular, for
$\alpha=\frac{1}{9}$, we retrieve the first model of chaotic inflation
in SUGRA proposed in 1983, which is known as the Goncharov-Linde
(GL) model, and it is well consistent with the present data \cite{Linde:2014hfa,Goncharov:1983mw,
Goncharov:1985yu}.
CAM were embedded in $ \mathcal{N}=1 $ SUGRA using superconformal symmetries by introducing a 3 chiral super multiplets: a conformon $ X^{0} $, an inflaton $ X^{1}=\Phi $ and a sGoldstino $X^{3}=S$ \cite{Kallosh:2013yoa,Kallosh:2013lkr,
Kallosh:2013pby}. In this set up single field inflation is achieved at the minimum of the superpotential by the requirement that the fields $S$ and Im$\Phi$ remain heavy during inflation\footnote{This mechanism has
also envisaged the multifield inflation with a curvaton, i.e, where
we can have generation of isocurvature perturbations when $S$ or
Im$\Phi$ are light and play the role of curvaton during or after
the end of inflation \cite{Kallosh:2010ug,Kallosh:2010xz,
Demozzi:2010aj}}. 
In recent studies, $ \alpha- $ attractors realized in SUGRA\footnote{Obtaining inflation from SUGRA also brings other benefits such as, exploring supersymmetry (SUSY) breaking sector
and the presence of dark energy \cite{Abe:2014opa,Linde:2014ela,Kallosh:2015lwa,
Carrasco:2015pla,Scalisi:2015qga}.}
by only requiring a single chiral superfield  \cite{Roest:2015qya,Linde:2015uga}. A generalization
of K\"ahler potentials for viable single field models with respect to
Planck data, plus their connection to open and closed string sector has been
investigated in \cite{Roest:2013aoa}. 

In this paper we study non-slow-roll inflaton dynamics in the $\alpha-$attractor
model using the recently proposed approach of Gong and Sasaki (GS)
\cite{Gong:2015ypa}, which constitutes, to our knowledge, a new strategy. 
More concretely, we focus on the
non-canonical aspect of the $\alpha-$ attractor model. We start with the assumption
of GS \cite{Gong:2015ypa}, where the number of e-foldings $N$
which is counted backward in time is assumed to be a function of the inflaton
field $\phi$ during inflation. We retrieve the local shape of the
potential during inflation which can be steep and allowing for $60$ e-foldings to occur.
More precisely, we restrict our study to the region of the potential
where inflation is occurring. We emphasize that both the pre- and post-inflationary dynamics are beyond the scope of this paper. Afterwards,
we explore the GS parametrization within our chosen inflaton dynamics showing that inflation occurs for a wider class of potentials. We further show that we can maintain the predictions of the $\alpha-$attractor model displayed in \cite{Kallosh:2013yoa}, but now herein retrieved alternatively within a non-slow-roll. Finally, we study 
the possibility of realizing this model within $\mathcal{N}=1$ SUGRA. We explore the relation between the inflaton dynamics and the corresponding K\"ahler geometry curvature. We also comment on the stability of inflaton  trajectory during inflation. 

The paper is organized as follows: In section \ref{alphaattractor},
we revise the $\alpha-$attractor model and present arguments
supporting a non-slow-roll approach for these models. In section \ref{non-slow-roll-dynamics},
we describe GS parametrization and implement the non-slow-roll dynamics
in the context of $\alpha-$attractors. In section \ref{PW},
we present predictions for a specific
case of the GS parametrization. In section \ref{Largesmallattractors},
we complement the previous predictions for a wider class of non-slow-roll
dynamics and discuss on large and small field
inflation. We show that these scenarios exhibit an attractor in the $\left(n_{s},\, r\right)$
plane and discuss the (dis)similarities with standard slow-roll inflaton
dynamics. In section \ref{SUGRAembedding}, we review the SUGRA realization
of this scenario and verify the stabilization of the inflaton trajectory
during inflation.

\section{$\alpha-$attractor model}

\label{alphaattractor} In this section, we revise the essentials
of $\alpha-$attractor models which have been studied under slow-roll frameworks so far as in  \cite{Galante:2014ifa,Kallosh:2013yoa,
Kallosh:2015lwa}
and provide a baseline for our interest on these models which we will
be exploring in the rest of the manuscript from a new perspective and methodology.

The Lagrangian for $\alpha-$attractor models, in the Einstein frame, is given by%
\footnote{We assume the units $M_{\textrm{Pl}}=1$.%
} \cite{Kallosh:2015lwa}

\begin{equation}
\mathcal{\mathcal{L}}_{E}=\sqrt{-g}\left[\frac{R}{2}-\frac{1}{\left(1-\phi^{2}/6\alpha\right)^{2}}\frac{\left(\partial\phi\right)^{2}}{2}-f^{2}\left(\phi/\sqrt{6\alpha}\right)\right]\,,\label{alphaL}
\end{equation}
where $\alpha=1$ leads to the same prediction of the Starobinsky
model (in the Einstein frame), $\alpha=1/9$ corresponds to GL model \cite{Linde:2014hfa},
and for large $\alpha$ this model is equivalent to chaotic inflation
with quadratic potential \cite{Linde:1983gd}. In order to prevent
 negative gravity in the Jordan frame it is required to have $\vert\phi\vert<\sqrt{6\alpha}$
\cite{Kallosh:2013yoa,Kallosh:2014rga}. Furthermore, in the SUGRA
embedding of this model, the parameter $\alpha$ is shown to be related
to the curvature of K\"ahler manifold as

\begin{equation}
\mathcal{R}_{\mathcal{K}}=-\frac{2}{3\alpha}\,.\label{kalhercurvature}
\end{equation}

The Lagrangian Eq.(\ref{alphaL}) is a subclass of $ k $-inflationary
model where the kinetic term is linear%
\footnote{$K\left(\phi\right)=1$ gives the canonical kinetic term.%
} in $X$, i.e.,

\begin{equation}
P\left(X,\phi\right)=K\left(\phi\right)X-f^{2}\left(\phi/\sqrt{6\alpha}\right)\,,\label{kinflationE}
\end{equation}
where $K\left(\phi\right)=\frac{1}{\left(1-\phi^{2}/6\alpha\right)^{2}}$
and $X=-\frac{\left(\partial\phi\right)^{2}}{2}$. The speed of sound
for these class of models is $c_{s}^{2}=1$ \cite{PDM99}, therefore
these models are not expected to show large non-Gaussianities \cite{Chen:2006nt}.

In this theory, the Friedmann equation is

\begin{equation}
H^{2}=\frac{1}{3}\left[XK\left(\phi\right)+f^{2}\left(\frac{\phi}{\sqrt{6\alpha}}\right)\right]\,.\label{Efriedmann}
\end{equation}
The Raychaudhuri equation is

\begin{equation}
\dot{H}=-XP_{,X}\,\:\: \text{with}\:\: P_{,X}= \frac{\partial P}{\partial X},\label{RaychaudhuriE}
\end{equation}

and the equation of motion for the scalar field is given by

\begin{equation}
\frac{d}{dt}\left(K\left(\phi\right)\dot{\phi}\right)+3HK\left(\phi\right)\dot{\phi}-P_{,\phi}=0\,.\label{Eom}
\end{equation}
In the literature it is found that inflation in the $\alpha-$attractor
model has been realized in terms of a canonically normalized field $\left(\varphi\right)$ as

\begin{equation}
\frac{d\varphi}{d\phi}=\frac{1}{\left(1-\frac{\phi^{2}}{6\alpha}\right)}\Rightarrow\frac{\phi}{\sqrt{6\alpha}}=\tanh\frac{\varphi}{\sqrt{6\alpha}}\,.\label{canonical field}
\end{equation}
In this case, flat potentials are natural and subsequent slow-roll
dynamics of $\varphi$ lead to viable inflationary scenario with respect
to the observational data. The predictions of $\left(n_{s},\, r\right)$
for these models are shown to be solely determined by the order and
residue of the Laurent series expansion leading pole in the kinetic
term \cite{Galante:2014ifa}. The slow-roll inflationary predictions
of $\alpha-$attractor models are

\begin{equation}
n_{s}=1-\frac{2}{N}\quad r=\frac{12\alpha}{N^{2}}\,.\label{sweetspot-1}
\end{equation}
In terms of this canonically normalized field $\left(\varphi\right)$ the
equation of motion (\ref{Eom}) becomes

\begin{equation}
\ddot{\varphi}+3H\dot{\varphi}+V_{,\varphi}=0\,.\label{canonicalEOM}
\end{equation}
Therefore, under slow-roll assumption this reduces to

\begin{equation}
3H\dot{\varphi}\simeq V_{,\varphi}\,.\label{slowroll}
\end{equation}

Our purpose is to obtain viable inflationary
predictions, by means of extending $ \alpha- $ attractors towards non-slow-roll dynamics. Therefore, in the present work, we restrict
ourselves to the range $\phi^{2}<6\alpha$. We will emphasize similarities
and of course the differences with the (canonically normalized
field) slow-roll inflation case. In the following section we unveil the context of non-slow-roll towards $\alpha-$attractors.

\section{Non-slow-roll dynamics }

\label{non-slow-roll-dynamics} 
The recent work by Gong \& Sasaki (GS)
\cite{Gong:2015ypa} points out a cautionary remark on applying slow-roll approximation
in the context of k-inflation. The argument, presented there, lies
in the fact that the second derivative term in the equation of motion
(\ref{Eom}) may not be negligible in general. In this regard,
the authors introduce a new parameter
\begin{equation}
p=\frac{\dot{P}_{,X}}{HP_{,X}}\,,
\end{equation}
which could bring significant differences in the local non-Gaussianity.
They have illustrated the role of this new parameter and observationally
viable inflationary scenarios in the context of some non-trivial examples. 

Let us implement the aforementioned procedure here in the context
of $\alpha-$attractors as $\phi$ is a non-canonical scalar field
given by Eq. (\ref{kinflationE}). This new approach enable us to
study the $\alpha-$attractors in the context of non-slow-roll by assuming
that the inflaton field during inflation behaves as%
\footnote{We start with a similar parametrization as the one used in section
3.2 of \cite{Gong:2015ypa}.%
}

\begin{equation}
\phi=n\exp\left(\beta N\right)\,,\label{sasakiparametrization}
\end{equation}
where $N=\ln a\left(t\right)$ is the number of efoldings counted
backward in time from the end of inflation and $n$ is treated as
a free parameter that specifies the value of the field at $N\rightarrow0$.
We assign Eq. (\ref{sasakiparametrization}) as GS parametrization
for subsequent reference. This parametrization is particularly useful in the cases of non-canonical scalar field models, whereas in Refs. \cite{Martin:2012pe,Motohashi:2014ppa} a different parametrization was applied to the case of canonical scalar field inflation. We declare here that our study of inflation in $ \alpha-$ attractor model is based on the dynamics for the inflaton assumed in Eq. (\ref{sasakiparametrization}) parametrized by $ \left(n,\,\beta\right) $. Therefore, we label our approach for the $ \alpha- $attractor framework as non-slow-roll, following the same terminology used in Ref. \cite{Gong:2015ypa}. Being more precise, in this paper we do not impose any slow-roll approximation in particular. We note at this point that non-slow-roll does not mean a non-smallness of conventional parameters $ \epsilon,\,\eta $ (see Ref. \cite{Gong:2015ypa} for more details). Moreover, and we stress that this is a most important point in our
study, we completely relax the choice of the inflaton potential and rather concentrate on the inflaton dynamics that can give rise to viable observational predictions.

Substituting $\phi$ from Eq. (\ref{sasakiparametrization})
in the Raychaudhuri equation we obtain

\begin{equation}
H^{\prime}=\frac{\alpha^{2}H\left(N\right)}{2}\phi^{2}K\left(\phi\right)\,,\label{rayintegrate}
\end{equation}
where the prime  $ ^{\prime} $ denotes differentiation with respect to $N$. Integrating
Eq. (\ref{rayintegrate}), we get

\begin{equation}
H\left(N\right)=\lambda e^{-\frac{9\beta\alpha^{2}}{\phi^{2}-6\alpha}}\,,\label{HEsol}
\end{equation}
where $\lambda$ is the integration constant. At this point, we should mention that our calculations are similar to the Hamilton-Jacobi like formalism found in \cite{Muslimov:1990be,Salopek:1990jq,Motohashi:2014ppa}.

Inserting the aforementioned
solution in the Friedmann Eq. (\ref{Efriedmann}), we can express the
local shape of the potential during inflation as

\begin{equation}
f^{2}\left(\frac{\phi}{\sqrt{6\alpha}}\right)=\lambda \exp\left({-\frac{18\beta\alpha^{2}}{\phi^{2}-6\alpha}}\right)\left[3-\frac{\beta^{2}\phi^{2}}{2\left(1-\frac{\phi^{2}}{6\alpha}\right)^{2}}\right]\,.\label{potential}
\end{equation}

It should therefore be noted that the suitable choice of potentials considered in the case of slow-roll $ \alpha- $attractors are quite different, namely, power law type $ V\sim\phi^{2n} $ in terms of original scalar field (or) T-models, i.e.,$  \,V\sim\tanh^{2n}{\frac{\varphi}{\sqrt{6\alpha}}} $ in terms of canonically normalized field \cite{Galante:2014ifa,Kallosh:2013yoa,
Kallosh:2015lwa}. In Ref. \cite{Kallosh:2014rga} the power law potentials were generalized to the following form of power series

\begin{equation}
f^{2}\left(\frac{\phi}{\sqrt{6\alpha}}\right)= \sum_{n} c_{n}\phi^{n}\,,
\label{seriespot}
\end{equation}
where $ c_{n} $ are non-zero constants and it was argued to be $ c_{0}\ll1 $. In this class of potentials the inflaton slow-rolls towards the potential minimum\footnote{It has been studied in the Ref. \cite{Cespedes:2015jga} that the slow-roll inflation in T-models may be interrupted abruptly in some cases of matter couplings to inflaton field.} which is located at $ \phi=0 $.

In the subsequent sections, with the assumed GS parametrization, we will show that non-slow-roll inflation occurs to be near the pole of the kinetic term i.e., $\vert\phi\vert\to\sqrt{6\alpha} $. Therefore, we can observe from Eq. (\ref{potential}) that the local shape of the potential in the non-slow-roll approach is different from the power-law (or) T-models and also the power series form given in Eq. (\ref{seriespot}). In this regard, our study about the non-slow-roll approach widens the scope for different shapes of inflationary potentials in $ \alpha- $ attractors.  

Subsequently, for the conventional parameters general definitions\footnote{The sign difference in the definition of parameters $ \epsilon,\,\eta $ is due
to $N$ which is counted backward in time from the end of inflation (see Eq. (\ref{epsilon})). %
}

\begin{equation}
\epsilon=\frac{H^{\prime}}{H}\quad,\quad\eta=-\frac{\epsilon^{\prime}}{\epsilon}\,,\label{epsilon}
\end{equation}
substituting the Hubble parameter from Eq. (\ref{HEsol}) and demanding the end of inflation $\epsilon=1$ at $N=0$
we get

\begin{equation}
\alpha=\frac{n^{2}}{3\sqrt{2}\beta n+6}\,.\label{alphafix}
\end{equation}
Consequently, constraining the parameter space $\left(n,\,\beta\right)$
automatically gives the values of $\alpha$. In the next sections we show that the $ \beta $ parameter determines the value of scalar spectral index $ n_{s} $, whereas as the parameter $ n $, which indicates the value of inflaton field at the end of inflation, regulates the tensor to scalar ratio $ r $. From Eqs. (\ref{sasakiparametrization}),
(\ref{potential}) and (\ref{alphafix}), we can say that the local
shape of the potential, the inflaton dynamics and the parameter $\alpha$
are interconnected. In other words, identifying $\alpha$ as the curvature
of K\"ahler geometry given by Eq. (\ref{kalhercurvature}), we can establish
a web of relations, 

\vspace{0.5cm}
\hspace{1cm}
\begin{picture}(0,0)%
   
    \put(65,0){\color[rgb]{0,0,0}\makebox(0,0)[lb]{K\"ahler Geometry}}%
    \put(165,0){\makebox(0,0)[lb]{{\Huge{$\leftrightharpoons$}}}}%
    \put(205,0){\color[rgb]{0,0,0}\makebox(0,0)[lb]{Inflaton Dynamics}}%
    
    \put(115,-15){\color[rgb]{0,0,0}\rotatebox{-45}{\makebox(0,0)[lb]
    {\Huge{$\leftrightharpoons$}}}}%
    \put(225,-5){\color[rgb]{0,0,0}\rotatebox{-135}{\makebox(0,0)[lb]
    {\Huge{$\leftrightharpoons$}}}}%
     \put(125,-45){\color[rgb]{0,0,0}\makebox(0,0)[lb]{Local shape of the potential}}%

\end{picture}%

\vspace{2cm}

From the above schematic diagram we can decipher that the class of potentials which are obtained by allowing different values for
$\left(n,\,\beta\right)$ is related to the family of K\"ahler geometries,
which determine the dynamics of inflaton during inflation. In the
next section, we derive the scalar and tensor power spectrum for this model. 

\section{Power spectrum}

\label{PW}

In this section, we derive the scalar and tensor spectral indices, the tensor to scalar ratio up to the third order in the parameters
$\epsilon\,,\,\eta$. We present our calculations, which are carried by assuming $ \eta\simeq $ constant. We closely follow the derivations presented in Refs. \cite{Kobayashi:2011nu,Khoury:2008wj}. Similar results can also be found in Refs. \cite{Ribeiro:2012ar,Zhu:2014wfa}.

\subsection{Scalar power spectrum}

The second order action for scalar perturbations in the $ k $-inflationary model with speed of sound $ c_{s}=1 $ is given by \cite{Kobayashi:2011nu,Khoury:2008wj}
\begin{equation}
S_{s}^{(2)}=\int\, dt\, d^{3}x\:\:\epsilon \: a^{3}\left[\dot{\zeta}^{2}-\dfrac{1}{a^{2}}\left(\nabla\zeta\right){}^{2}\right]\,,\label{S2pert}
\end{equation}
where $\zeta$ is the curvature perturbation, $t$ is the cosmic time. 

To quantify the amplitude and tilt of the spectrum we use the
variables\footnote{We are following a similar notation as the one used in Ref. \cite{Kobayashi:2011nu}} $dy_{s}\equiv\frac{1}{a}\, dt$, $z_{s}\equiv\sqrt{2\epsilon}\:a$ and $u\equiv z_{s}\zeta$. The action (\ref{S2pert}) can be
canonically normalized as 
\begin{equation}
S_{s}^{(2)}=\frac{1}{2}\int dy_{s}\, d^{3}x\left[(u')^{2}-(\nabla u)^{2}+\frac{z_{s}''}{z_{s}}u^{2}\right]\,\:,\label{SS2}
\end{equation}
where $^{\prime}$ denotes differentiation with respect to $ y_{s} $. Integrating by parts $dy_{s}=\left(1/a\right)\, dt$, by assuming that $\eta$ is sufficiently small and constant, we get (cf. \cite{Khoury:2008wj}) 
\begin{equation}
y_{s}=\frac{1}{\left(\epsilon-1\right)\, aH}\left(1+\frac{\epsilon\eta}{(\epsilon-1)^{2}}\right)\,.\label{ys}
\end{equation}
The equation of motion for the mode function $u_{k}$ is given by, 

\begin{equation}
y_{s}^{2}u_{k}^{\prime\prime}+y_{s}^{2}k^{2}u_{k}-\left(\nu_{s}^{2}-\frac{1}{4}\right)u_{k}=0\: ,\label{modeEqn}
\end{equation}
where 
\begin{equation}
\nu_{s}^{2}-\frac{1}{4}\equiv y_{s}^{2}\frac{z_{s}''}{z_{s}}\: .\label{nus}
\end{equation}
Imposing the flat spacetime vacuum solution in the subhorizon limit
$k/aH\to\infty$ for the perturbation mode $u_{k}$, we find
the solution for the Eq. (\ref{modeEqn}) as 
\begin{equation}
u_{k}=\frac{\sqrt{\pi}}{2}\sqrt{-y_{s}}\, H_{\nu_{s}}\left(-ky_{s}\right)\,.\label{scalarmodefuntion}
\end{equation}
Using now $\zeta=u/z_{s}$, we obtain the power spectrum of primordial curvature perturbation as

\begin{equation}
{\cal P}_{\zeta}=\frac{\gamma_{s}}{2}\frac{H^{2}}{4\pi^{2}\epsilon}\biggr\lvert_{k=aH}\quad,\quad\gamma_{s}\equiv2^{2\nu_{s}-3}\frac{\Gamma\left(\nu_{s}\right)^{2}}{\Gamma(3/2)^{2}}\left(1-\epsilon\right)^{2}\,.\label{pwrspectrum}
\end{equation} 
From Planck data \cite{Ade:2015lrj} the power spectrum amplitude
is known to be ${\cal P}_{\zeta_{*}}=2.2\times10^{-9}$. Using this
bound, with Eqs. (\ref{HEsol}) and (\ref{pwrspectrum}), we constrain
$\lambda\sim\mathcal{O}\left(10^{-6}\right)$.

Finally the scalar spectral index is given by

\begin{equation}
n_{s}-1=3-2\nu_{s}\,.\label{ns-1}
\end{equation}
Calculating $\nu_{s}$ up to the third order in the parameters $ \epsilon,\,\eta $,
by using the definition of $z_{s}$ and Eq. (\ref{ys}), we obtain 
\begin{equation}
\begin{aligned}\nu_{s}= & \left(\frac{3}{2}+\epsilon+\epsilon^{2}+\epsilon^{3}\right)+\left(\frac{1}{2}+2\epsilon+\frac{29\epsilon^{2}}{6}+\frac{82\epsilon^{3}}{9}\right)\eta+\left(-\frac{1}{6}+\frac{23\epsilon}{18}+\frac{1069\epsilon^{2}}{108}+\frac{5807\epsilon^{3}}{162}\right)\eta^{2}+\\
 & \left(\frac{1}{18}+\frac{23\epsilon}{54}+\frac{707\epsilon^{2}}{108}+\frac{19633\epsilon^{3}}{486}\right)\eta^{3}+\mathcal{O}\left(\epsilon^{4}\,,\,\eta^{4}\right)\,.
\end{aligned}
\label{nusfull}
\end{equation}

\subsection{Tensor Power spectrum}

The tensor power spectrum derivation follows closely
the scalar power spectrum one. The second order action for tensor
perturbations can be written as \cite{Kobayashi:2011nu,Khoury:2008wj}
\begin{equation}
S_{t}^{(2)}=\frac{1}{8}\int\, dt\, d^{3}xa^{3}\left[\dot{h}_{ij}^{2}-\dfrac{1}{a^{2}}\left(\nabla h_{ij}\right){}^{2}\right]\,,\label{st2}
\end{equation}
We use the variables $dy_{t}\equiv\frac{1}{a}\, dt$, $z_{t}\equiv\frac{a}{2}$
and $u_{ij}\equiv z_{t}h_{ij}$ so that the action (\ref{st2})
can be canonically normalized as 
\begin{equation}
S_{t}^{(2)}=\frac{1}{2}\int dy_{t}\, d^{3}x\left[(u_{ij}')^{2}-(\nabla u_{ij})^{2}+\frac{z_{t}''}{z_{t}}u_{ij}^{2}\right]\,.\label{modeteq}
\end{equation}
Integrating by parts $dy_{t}=\left(1/a\right)\, dt$, again by assuming that $\eta$
is sufficiently small and constant, we obtain
\begin{equation}
y_{t}=\frac{1}{\left(\epsilon-1\right)\, aH}\left(1+\frac{\epsilon\eta}{\left(\epsilon-1\right)^{2}}\right)\,.\label{yt}
\end{equation}
Imposing the flat spacetime vacuum solution as in Eq.~(\ref{scalarmodefuntion})
we find 
\begin{equation}
u_{ij}=\frac{\sqrt{\pi}}{2}\sqrt{-y_{t}}\, H_{\nu_{t}}^{(1)}\left(-ky_{t}\right)\, e_{ij}\,\quad,\quad\nu_{t}^{2}-\frac{1}{4}\equiv y_{t}^{2}\frac{z_{t}''}{z_{t}}\,.\label{tensor modefunction}
\end{equation}
where $e_{ij}$ is the polarization tensor. Recalling that $h_{ij}=u_{ij}/z_{t}$
and taking into account the two polarization states, we obtain the power spectrum of primordial tensor perturbation

\begin{equation}
{\cal P}_{T}=8\gamma_{t}\frac{H^{2}}{4\pi^{2}}\biggr\lvert_{k=aH}\quad,\quad\gamma_{t}\equiv2^{2\nu_{t}-3}\frac{\Gamma\left(\nu_{t}\right)^{2}}{\Gamma(3/2)^{2}}\left(1-\epsilon\right)^{2}\,.\label{ptspectrum}
\end{equation}
The tensor tilt $\left(n_{t}\right)$ is given by 
\begin{equation}
n_{t}=3-2\nu_{t}\,.\label{nt}
\end{equation}
Calculating $\nu_{t}$ up to the third order in the parameters $ \epsilon,\,\eta $,
by using the definition of $z_{t}$ and Eq. (\ref{ys}), we obtain 
\begin{equation}
\begin{aligned}\nu_{t}= & \left(\frac{3}{2}+\epsilon+\epsilon^{2}+\epsilon^{3}\right)+\left(\frac{4\epsilon}{3}+\frac{37\epsilon^{2}}{9}+\frac{226\epsilon^{3}}{27}\right)\eta+\left(\epsilon+\frac{227\epsilon^{2}}{27}+\frac{875\epsilon^{3}}{27}\right)\eta^{2}+\\
 & \left(\frac{28\epsilon^{2}}{9}+\frac{6491\epsilon^{3}}{243}\right)\eta^{3}+\mathcal{O}\left(\epsilon^{4}\,,\,\eta^{4}\right)\,.
\end{aligned}
\label{nut}
\end{equation}
From Eqs.~(\ref{pwrspectrum}) and (\ref{ptspectrum}) we define the tensor to scalar ratio as

\begin{equation}
r=\dfrac{\cal P_{T}}{\cal P_{\zeta}}=16\dfrac{\gamma_{t}}{\gamma_{s}}\epsilon\,.\label{TSR}
\end{equation}


\subsection{Inflationary predictions for $n=1$}

\label{inflaitionary predictionsn1}

In this section, we study the inflationary predictions of the model taking $n=1$. We constrain the parameter $\beta$ to obtain the predictions of $\left(n_{s}\,,\,r\right)$  within current observational range. 

\begin{figure}[t]
\centering\includegraphics[height=2.5in]{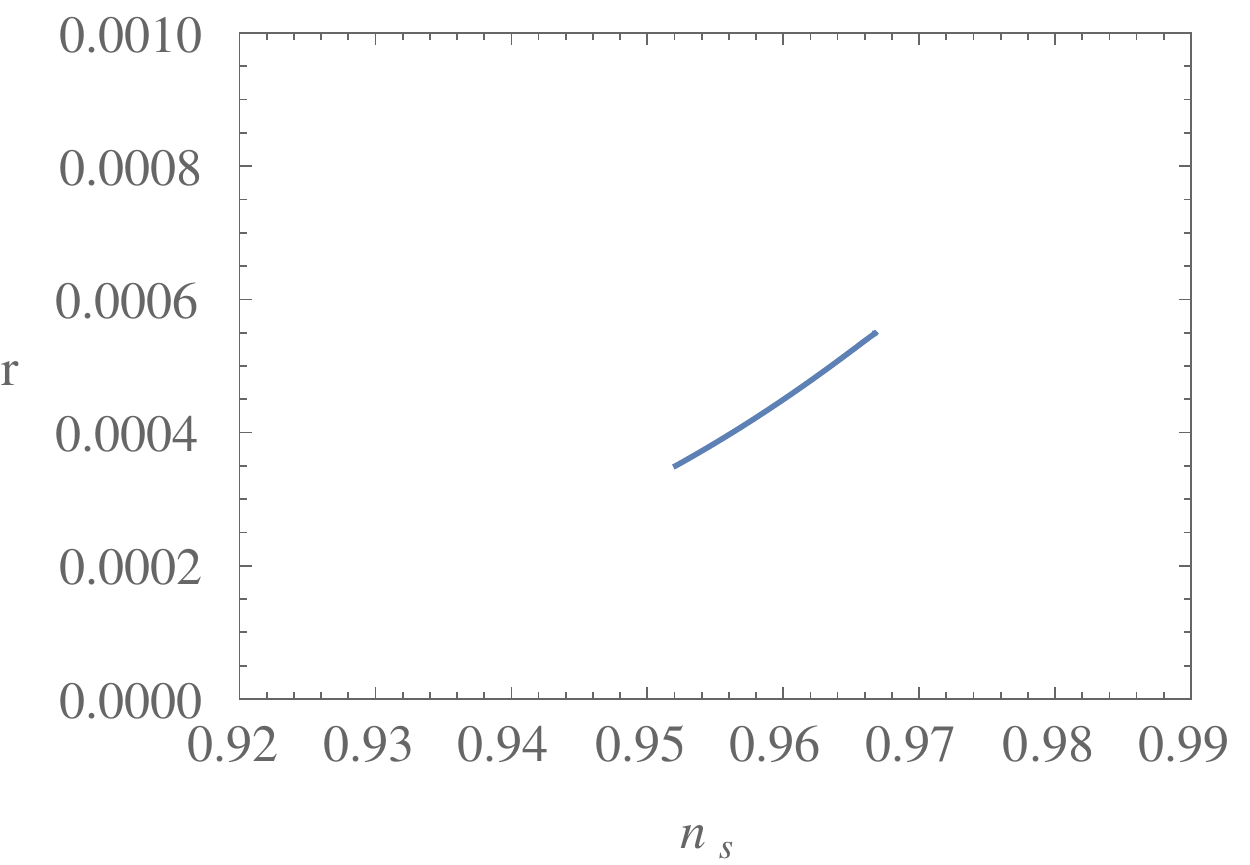}\quad{}\caption{Parametric plot of spectral index $\left(n_{s}\right)$ verses tensor
scalar ratio $\left(r\right)$. We have considered 60 number of efoldings with $n=1\,,\,-0.03<\beta<-0.001$ (or equivalently $0.166\lesssim\alpha\lesssim0.17$).}
\label{ns-r-nt-t} 
\end{figure}

Imposing the spectral index $n_{s}=0.968\pm0.006$, we obtain the constraint
$\vert\beta\vert\sim\mathcal{O}\left(10^{-3}\right)$ (or equivalently,
from Eq. (\ref{alphafix}), $\alpha\sim\mathcal{O}\left(10^{-1}\right)$).
However, we verify that the inflaton dynamics for the case $\beta>0$
violates the requirement that $\phi^{2}<6\alpha$. Therefore, we only
consider the case with $\beta<0$ as a viable inflationary paradigm
complying with $\phi^{2}<6\alpha$ during inflation. In this case,
we find that inflation occurs while approaching asymptotically the
kinetic term pole at $\vert\phi\vert\rightarrow\sqrt{6\alpha}$. The
predictions of $\left(n_{s},\, r\right)$ are depicted in the Fig. \ref{ns-r-nt-t}. 

\begin{figure}[t]
\centering\includegraphics[height=2in]{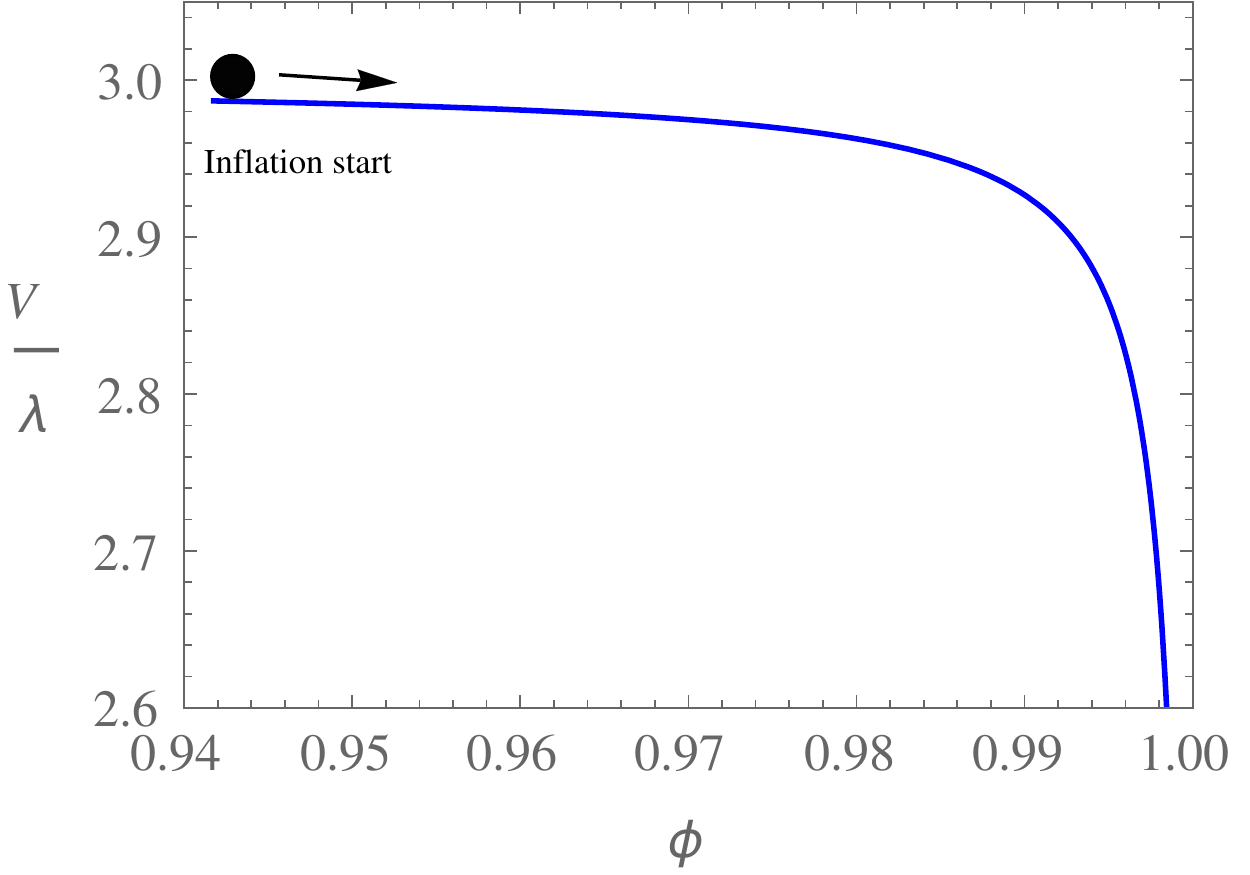}\quad{}\includegraphics[height=2in]{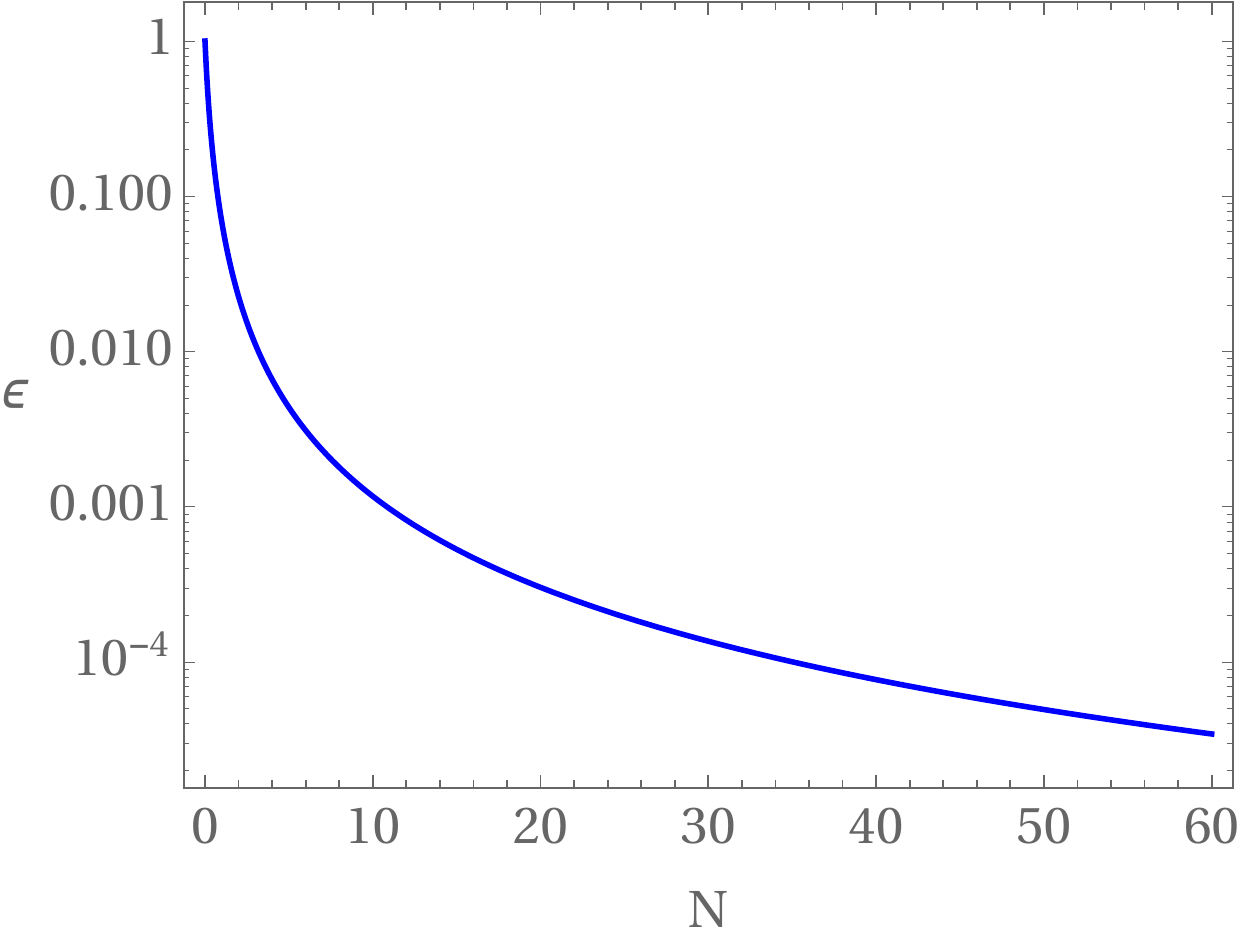}\quad{}\caption{The left panel is the graphical presentation of the local shape of the potential verses scalar field during inflation. The right panel
depicts the parameter $\epsilon\;$verses$\;N$. We have taken $\beta=-0.001$ (or equivalently $\alpha=0.167$) for both plots.}

\label{pot-slow} 
\end{figure}

The left panel of Fig. \ref{pot-slow} depict
the shape of the potential during which inflation is happening in the non-slow-roll context. 
In the right panel of Fig. \ref{pot-slow}, we plot the parameter $\epsilon\;$verses$\;N$ 
for a particular value of $\alpha$ corresponds to $ n=1 $. 

In addition, we compute the energy scale of inflation and mass of
the inflaton $\left(m_{\phi}^{2}\right)$ by computing the $V_{*}^{1/4}$
and the $\partial_{\phi}^{2}V_{*}$ where $V_{*}$ is the the potential
evaluated at horizon exit. In this context, the shape of the potential
during inflation is given by Eq. (\ref{potential}), consequently
we obtain,

\begin{equation}
f_{*}^{1/2}\sim1.2\times10^{17}\:\textrm{GeV}\quad,\quad m_{\phi}^{2}<0\,.\label{energymass}
\end{equation}

Therefore, since the energy scale of inflation appears to be greater than
GUT scale but still below Planck scale, this naturally justify the embedding
of this model in SUGRA. Since the mass squared of the inflaton is
negative, inflation is driven by a tachyonic field.

\section{Non-slow-roll $\alpha-$attractor}

\label{Largesmallattractors}In section \ref{inflaitionary predictionsn1},
we have studied non-slow-roll inflation with GS parametrization and
$n=1$, in this case we obtained $r\sim\mathcal{O}\left(10^{-4}\right)$.
The objective, at this point, is to assess inflationary scenarios
with any value of $r<0.1$, by allowing $n\neq1$ in Eq. (\ref{alphafix}). 

\subsection{Conditions for small field and large field inflation}

In this section, we study the parameter space of the model allowing the inflaton to do large and small
field excursions during inflation. We address the possibility of large and small field inflation in the context of non-slow-roll
dynamics in $\alpha-$attractors.

\begin{figure}[t]
\centering\includegraphics[height=2.4in]{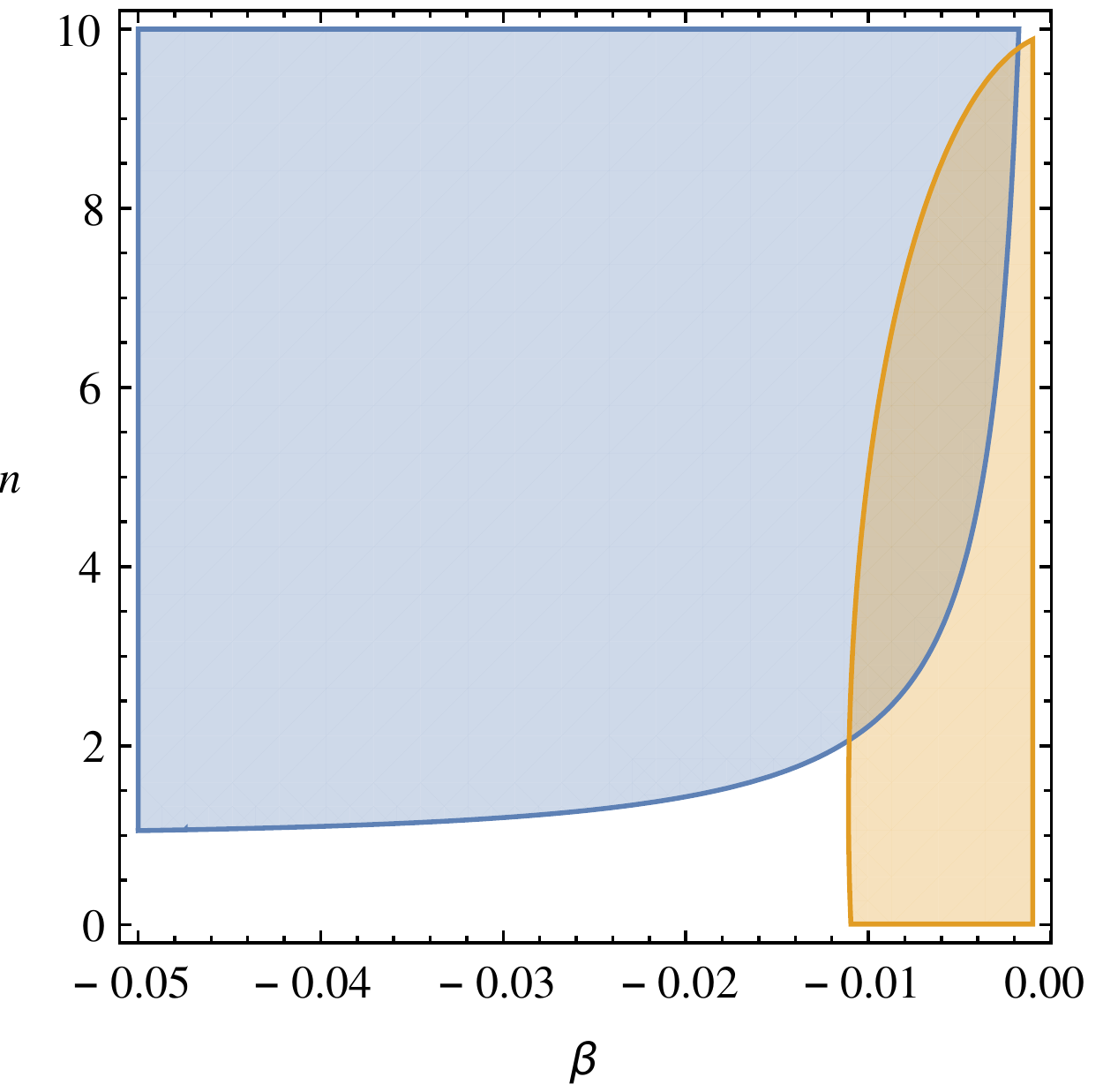}\quad{}\includegraphics[height=2.4in]{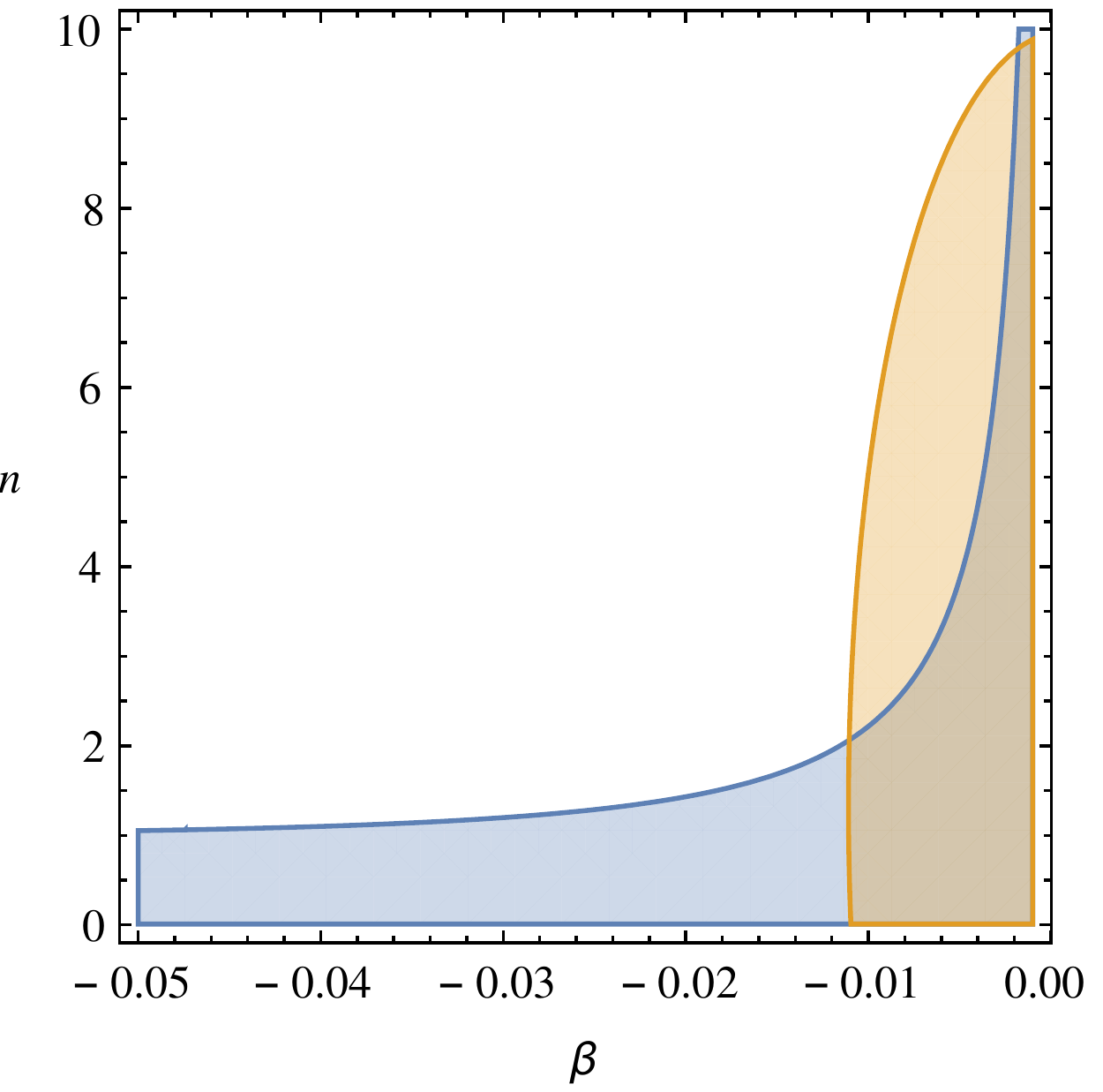}\quad{}\caption{In both plots the overlap of the blue and orange shaded regions satisfies the constraint $ 0.962<n_{s}<0.974 $. The blue shaded region in the left panel is for large field $ \Delta\phi>1 $ whereas in the right panel is for small field $ \Delta\phi<1 $.
We have considered $N=60$.}

\label{nalphalargesmall} 
\end{figure}

Using the parametrization from Eq. (\ref{sasakiparametrization})
the field excursion during the period of inflation is given by

\begin{equation}
\Delta\phi=n\left(1-\exp\left(60\beta\right)\right)\,.\label{fieldexcursion}
\end{equation}
The above relation allows us to identify the parameter space of $\left(n,\,\beta\right)$
to explicit the region of large field $\left(\Delta\phi>1\right)$ and
small field $\left(\Delta\phi<1\right)$ inflation (see Fig. \ref{nalphalargesmall}).
We further constrain the parameter space using Eq. (\ref{ns-1}), by imposing $0.962<n_{s}<0.974$ which is the $95\%$ CL region
given by Planck 2015. This constraint on spectral index confine $-0.001<\beta<-0.01$,
and precisely $\beta\sim-0.002$ corresponds to the central value of $n_{s}\sim0.967$.

\begin{figure}[t]
\centering\includegraphics[height=2in]{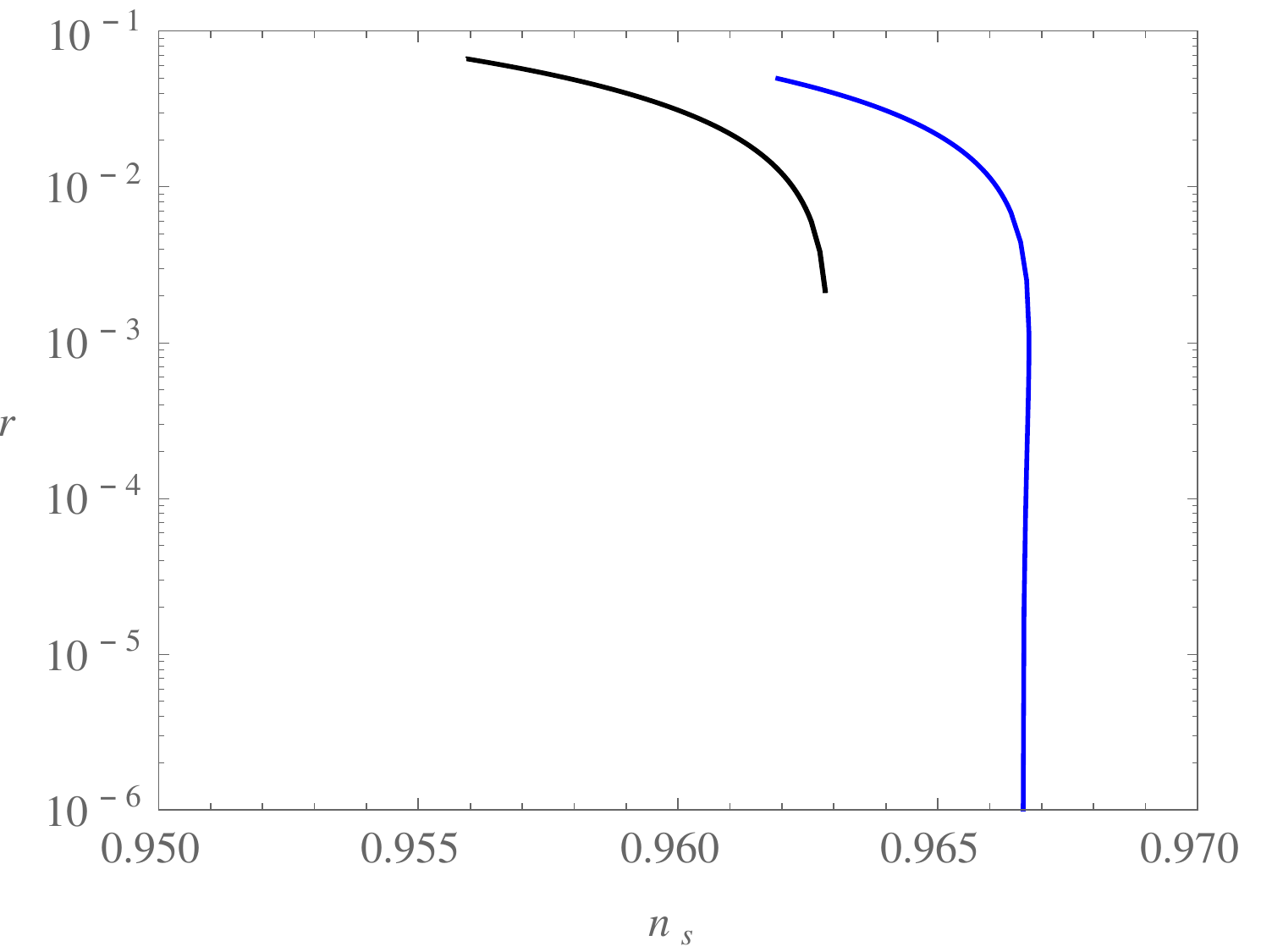}\quad{}\includegraphics[height=2in]{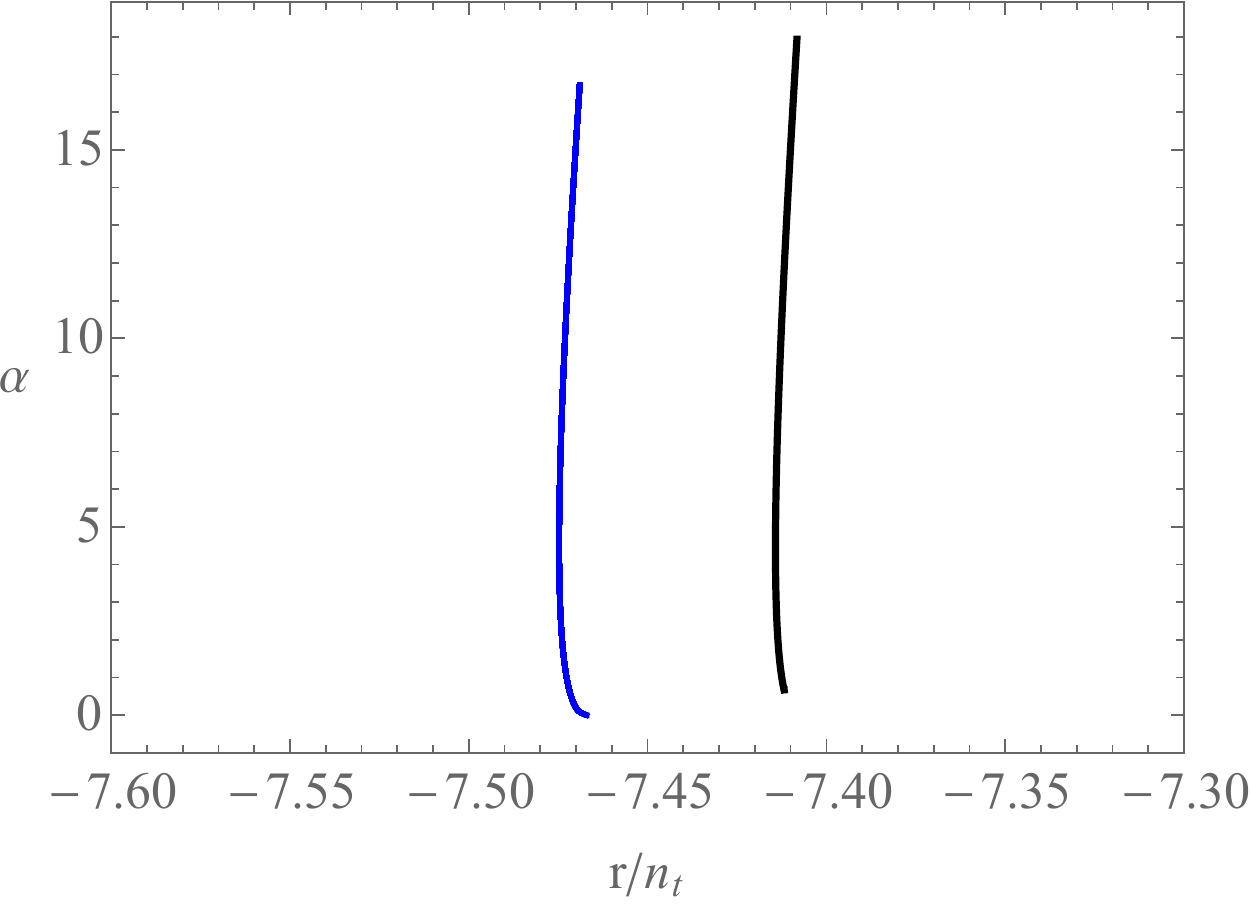}\quad{}\caption{Parametric plots of spectral index $\left(n_{s}\right)$ verses tensor
scalar ratio $\left(r\right)$ (left panel),  $\alpha$
verses the ratio of tensor scalar ratio and tensor tilt (right panel). In these plots the blue line
denote predictions for small field inflation for which we take $\beta\sim-0.002$
and $0<n<10$. In this case $r\rightarrow0$ as $n\rightarrow0$ (equivalently
$\alpha\rightarrow0$). The black line denote predictions for large
field inflation for which $\beta\sim-0.01$ and $2<n<10$. In this
case $r\gtrsim\mathcal{O}\left(10^{-3}\right)$. We have considered
$N=60$. }

\label{large-small-nsrntr} 
\end{figure}

The relation between tensor to scalar ratio and field excursion
during the period of inflation is defined by Lyth bound \cite{Lyth:1996im} which is 

\begin{equation}
\Delta\phi>\sqrt{\frac{r}{8}}N_{e}\,,\label{lyth bound}
\end{equation}
where $N_{e}=60$ which is the number of e-foldings before the
end of inflation. We can see from the above relation that $r>0.002$
implies $\Delta\phi>\textrm{M}_{\textrm{Pl}}$, i.e, large field inflation. However, this bound gets modified for the $ k $-inflationary models \cite{Baumann:2006cd}. In this case, the generalization of Eq. (\ref{lyth bound}) is given by

\begin{equation}
\Delta\phi>\int_{0}^{N_{e}}\sqrt{\frac{r}{8}\:\frac{1}{c_{s}\:P_{,X}}}dN\,.\label{newlyth}
\end{equation}
where the sound speed $ c_{s}=1 $ in the case of $ \alpha- $ attractor model. In Eq. (\ref{newlyth}) the term $ P_{,X}=\left(1-\frac{\phi^{2}}{6\alpha}\right)^{-2} $ affects Lyth bound depending on the value of the parameter $ \alpha $. From Eq. (\ref{alphafix}) we know that the $ \alpha $ parameter is directly related to the inflaton dynamics. In Fig. \ref{nalphalargesmall}, we depict the parameter space for large and small field inflation overlapped on the  region where $ 0.962<n_{s}<0.974 $. Here, we explicitly characterize the possibility of super planckian excursion of the field $ \phi $ attributing to the field value at the end of inflation $ n\gtrsim2 $ and the parameter $ \beta\sim-0.01 $ (see left panel of Fig. \ref{nalphalargesmall}). The field $ \phi $ is sub planckian for  $ 0<n<\mathcal{O}\left(10\right) $ and the parameter $ \beta\sim-0.002 $ (see right panel of Fig. \ref{nalphalargesmall}). We present the corresponding predictions in Fig. \ref{large-small-nsrntr}, where we found that the large field inflation in the non-slow-roll context can give rise to the tensor to scalar ratio $0.003\lesssim r<0.1$ and the spectral index $ 0.955\lesssim n_{s}\lesssim 0.964 $. Whereas in the case of small field we obtain $0\lesssim r<0.1$ and the spectral index $ 0.96\lesssim n_{s}\lesssim 0.967 $.

\begin{figure}[t]
\centering\includegraphics[height=2.2in]{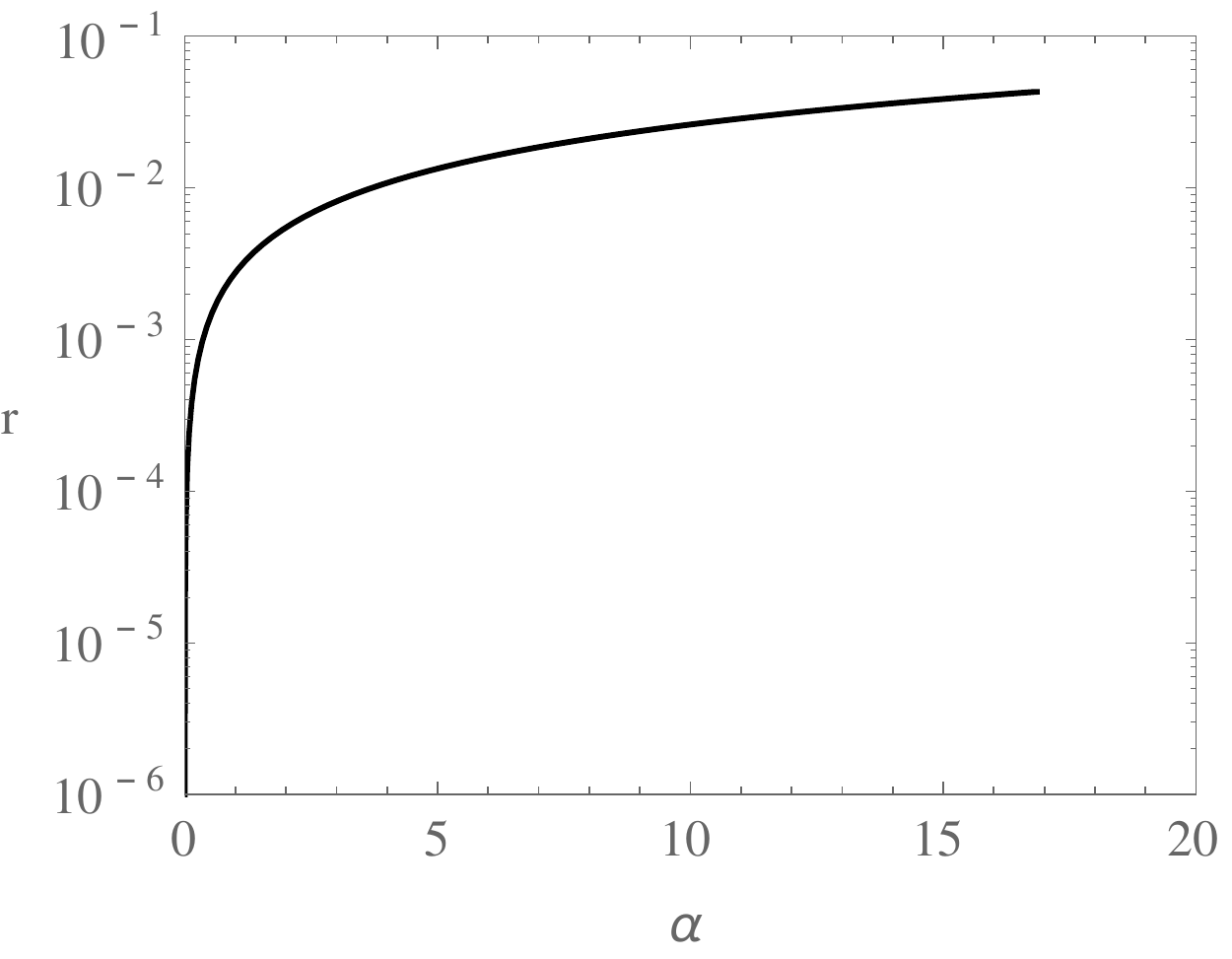}\quad{}\caption{Plot of tensor scalar ratio $\left(r\right)$ verses $\alpha$. Here we
have taken $\beta\sim-0.002$ and $0<n<10$. This plot is for $N=60$.}

\label{rattractor} 
\end{figure}

The parametrization used in Eq. (\ref{sasakiparametrization}) leads
to an attractor starting at $r\sim5.5\times10^{-4}$ which is the
prediction for $n=1$. We find that $r\rightarrow0$ as $n\rightarrow0$
(or equivalently $\alpha\rightarrow0$). We depict this behavior in
Fig. \ref{rattractor}. This attractor behaviour resembles with the
recently studied E-models \cite{Carrasco:2015rva}. The most interesting feature of our study is that, even with non-slow-roll dynamics of the inflaton, $\alpha-$attractors still appear to be the most promising models in the $\left(n_{s},\, r\right)$ plane. Including the higher order corrections in Eqs. (\ref{nusfull}) and (\ref{nut}) we have undetectably small deviation from the standard consistency relation $r=-8n_{t}$ as presented in the right panel of Fig. \ref{large-small-nsrntr}. However, the validity of the standard consistency relation remains
an open question and not even expected to be tested in any future CMB observations \cite{Errard:2015cxa}.

\section{Embedding in $\mathcal{N}=1$ SUGRA}

\label{SUGRAembedding}In this section, we revise the embedding of
$\alpha-$attractor within $\mathcal{N}=1$ SUGRA \cite{Kallosh:2013yoa}
and verify the stability of inflaton trajectory \cite{Kallosh:2010xz,Kallosh:2010ug}
in the context of non-slow-roll dynamics. 

The $\alpha-$attractor model can be embedded in SUGRA using 3 chiral
multiplets: a conformon $X^{0}$, an inflaton $X^{1}=\Phi=\frac{\phi+i\sigma}{\sqrt{2}}$
and a sGoldstino $X^{2}=S$. In order to extract a Poincar\'e SUGRA
conformon is gauge fixed as $X^{0}=\overline{X^{0}}=\sqrt{3}$. We
write the K\"ahler and superpotential in the similar way as studied
in Refs. \cite{Kallosh:2013yoa,Kallosh:2015lwa},

\begin{equation}
\mathcal{K}=-3\alpha\log\left(1-Z\overline{Z}-\frac{S\overline{S}}{3\alpha}+\frac{g}{3\alpha^{2}}\frac{\left(S\overline{S}\right)^{2}}{\left(1-Z\overline{Z}\right)}-\frac{\gamma}{3\alpha^{2}}\frac{S\overline{S}\left(Z-\overline{Z}\right)^{2}}{\left(1-Z\overline{Z}\right)^{2}}\right)\,,\label{kalher}
\end{equation}

\begin{equation}
W=Sf\left(Z\right)\left(1-Z^{2}\right)^{\left(3\alpha-1\right)/2}\,,\label{superpotential}
\end{equation}
where $Z=\frac{X^{1}}{X^{0}}=\frac{\Phi}{\sqrt{6\alpha}}$ and $f\left(Z\right)$
is an arbitrary function and the square of which serves as the inflaton
potential along $S=$Im$\Phi=0$. In the K\"ahler potential in Eq. (\ref{kalher})
we added an extra term $\frac{S\overline{S}\left(Z-\overline{Z}\right)^{2}}{\left(1-Z\overline{Z}\right)^{2}}$
in order to stabilize the inflaton trajectory in the direction of
$\textrm{Im}\Phi$ for any value of $\alpha$.
Although in some cases it is not required to add this extra term \cite{Kallosh:2015lwa,Carrasco:2015rva}.
In our case, we only focus our attention to the form of K\"ahler potential
given by Eq. (\ref{kalher}). 

The mass squares of $S$ and Im$\Phi$ for a given K\"ahler potential
are given by \cite{Kallosh:2010xz},

\begin{equation}
\begin{aligned}m_{\sigma}^{2} & =2\left(1-\mathcal{K}_{\Phi\overline{\Phi}S\overline{S}}\right)f^{2}+\left(\partial_{\Phi}f\right)^{2}-f\partial_{\Phi}^{2}f\\
m_{s}^{2} & =-\mathcal{K}_{S\overline{S}S\overline{S}}f^{2}+\left(\partial_{\Phi}f\right)^{2}\,,
\end{aligned}
\,,\label{supermasses}
\end{equation}
where all the terms in Eq. (\ref{supermasses}) are to be evaluated
along the inflaton trajectory $S=\textrm{Im\ensuremath{\Phi}=0. }$And
here $\mathcal{K}_{a\overline{b}c\overline{d}}=\partial_{a}\partial_{\overline{b}}\partial_{c}\partial_{\overline{d}}\mathcal{K}$.
For the stability of the inflaton trajectory it is required to have
$m_{\sigma}^{2}\,,\, m_{s}^{2}\gg H^{2}$ during inflation, in order
to ensure the absence of isocurvature perturbations and therefore
to have inflation solely driven by a single field \cite{Kallosh:2010xz}. 

For the K\"ahler potential given by Eq. (\ref{kalher}) we obtain,

\begin{equation}
\mathcal{K}_{\Phi\overline{\Phi}S\overline{S}}=-\frac{36\alpha^{2}\left(6\left(\alpha-2\gamma\right)+\phi^{2}\right)}{\left(\phi^{2}-6\alpha\right)^{3}}\qquad\mathcal{K}_{S\overline{S}S\overline{S}}=\frac{24\alpha(1-6g)}{\left(\phi^{2}-6\alpha\right)^{2}}\,.\label{kderivatives}
\end{equation}

\begin{figure}[t]
\centering\includegraphics[height=2.2in]{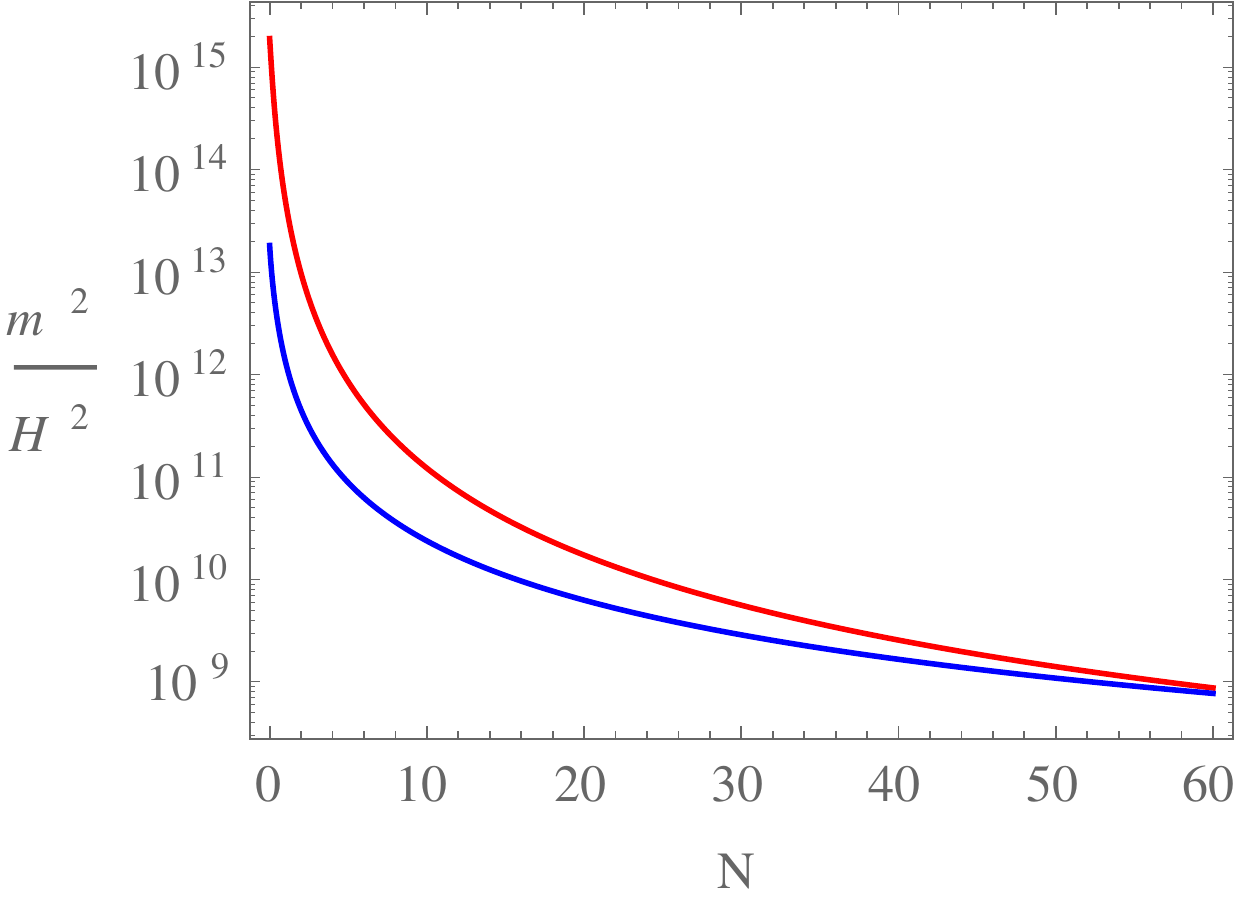}\quad{}\caption{In this figure we depict the ratio of the square of masses to the
square of Hubble parameter $H^{2}$. The red line indicates for Im$\Phi$
and the blue line is for $S$. We have taken $n=1$, $\alpha=0.167$,
$g=0.5$ and $\gamma=0.2$. }

\label{stability} 
\end{figure}

Evaluating the masses $m_{\sigma}^{2}$ and $m_{s}^{2}$ for the local
shape of inflaton potential given by Eq. (\ref{potential}) for $n=1$,
we obtain $m_{s}^{2},m_{\sigma}^{2}\gg H^{2}$ for $g\,,\,\gamma\geq0.2$
and for $\alpha\sim0.17$. For example, in Fig. \ref{stability},
we depict the ratio of inflaton mass square to Hubble parameter square
during inflation for a chosen values of $\left(g\,,\,\gamma\right)$. 

We can similarly verify the stability of the inflaton trajectory for
$n\neq1$ by appropriate choice of free parameters $\left(g\,,\,\gamma\right)$.

\section{Conclusions and Outlook}

In this work we have considered the $\alpha-$attractor models from a new perspective, more precisely, employing the framework of non-slow-roll approach in the way it was recently proposed
by Gong and Sasaki \cite{Gong:2015ypa}. We found that the $\alpha-$attractor models are quite compatible
in the $\left(n_{s},\, r\right)$ plane of Planck 2015 within non-slow-roll inflaton dynamics. We showed that such a particular inflationary scenario predicts an attractor at $n_{s}\approx0.967$ and $r\approx5.5\times10^{-4}$. We further found that the model can in principle predict any $r<0.1$. In addition,  we have extracted relation (\ref{alphafix}) between the $ \alpha- $ parameter, to the curvature of K\"ahler geometry, and to the inflaton dynamics. In other words, in our model, the curvature of the K\"ahler geometry defines the  local shape of the inflaton potential during inflation. This constitutes an interesting phenomenon which might be useful to understand the pre-inflationary physics. Furthermore, we also studied the possibility of large and small field inflation in the non-slow-roll context and contrasted them in terms of the predictions of the tensor to scalar ratio. 

In this work we have not considered any particular form for the inflaton potential, since the
assumption of $N=N\left(\phi\right)$ during inflation provides all
the necessary ingredients for studying inflation. It would be
interesting to look at the reheating phase in this model, given an adequate assumption about the shape of the potential in the post inflationnary epoch. In the view of the recent literature about the reheating process regarding Starobinsky inflation \cite{Terada:2014uia}, it would also  be interesting to consider the role of the superfields and their various possible decay modes, as well as of the inflaton. There are other possible directions to be explored, in the context of non-slow-roll dynamics in $\alpha-$attractors with respect to SUGRA, such as studying the various mechanisms for SUSY breaking and the origin of dark energy. In the sequence of this work, we are considering the extension of the non-slow-roll dynamics to the $\xi-$attractors case \cite{Kumaretal}.

\section{Acknowledgements}

We would like to thank the anonymous referee for useful comments.
SK is grateful to the IIT Kanpur for the hospitality and to the organizers
of the conference on "Primordial Universe after Planck" at the
Institut d'Astrophysique de Paris where part of this work was developed.
SK acknowledges for the support of grant SFRH/BD/51980/2012 from Portuguese
Agency Funda\c{c}\~ao para a Ci\^encia e Tecnologia. This research work is
supported by the grants PTDC/FIS/111032/2009 and UID/MAT/00212/2013.
Work of SD is supported by Department of Science and Technology, Government
of India under the Grant Agreement number IFA13-PH-77 (INSPIRE Faculty
Award).

\bibliographystyle{utphys}
\bibliography{References}

\end{document}